\documentclass[aps,prl,twocolumn,amsfonts,amssymb,amsmath,tightenlines,a4paper,notitlepage]{revtex4-1}
\pdfoutput=1
\usepackage{graphicx}
\usepackage{hyperref}
\usepackage[utf8]{inputenc}
\usepackage{ulem}

\newcommand{\dd}[0]{\mathrm{d}}

\newcommand{\matr}[1]{\mathbf{#1}}
\newcommand{\vect}[1]{\mathbf{#1}}
\newcommand{\transp}{\mathsf{T}}

\begin{document}


\title{Emergence of active nematic behaviour in monolayers of isotropic cells}

\author{Romain Mueller}
\author{Julia M. Yeomans}
\author{Amin Doostmohammadi}
\affiliation{The Rudolf Peierls Centre for Theoretical Physics, Clarendon Laboratory, Parks Road, University of Oxford, Oxford, OX1 3PU, UK}

\begin{abstract}
    There is now growing evidence of the emergence and biological functionality of liquid crystal features, including nematic order and topological defects, in cellular tissues.
    However, how such features that intrinsically rely on particle elongation, emerge in monolayers of cells with isotropic shapes is an outstanding question.
    In this article we present a minimal model of cellular monolayers based on cell deformation and force transmission at the cell-cell interface that explains the formation of topological defects and captures the flow-field and stress patterns around them.
    By including mechanical properties at the individual cell level, we further show that the instability that drives the formation of topological defects and leads to active turbulence, emerges from a feedback between shape deformation and active driving. The model allows us to suggest new explanations for experimental observations in tissue mechanics, and to propose designs for future experiments.
    
\end{abstract}

\maketitle


The collective migration of cells plays a crucial role in vital physiological processes, and there is rapidly growing interest in studying the interplay between mechanics and the collective behaviour of cells at the tissue level~\cite{doi:10.1038/nrm.2017.98}.
Experimental studies have uncovered the important role of mechanical forces in wound healing~\cite{doi:10.1038/nphys3040}, morphogenesis~\cite{Chiou17}, and cancer invasion~\cite{Friedl12}.
Interestingly, many cellular systems display properties of liquid crystals such as local nematic alignment and the appearance of topological defects --- singular points in the cellular alignment where the orientational order vanishes.
Important examples are elongated fibroblasts at high densities~\cite{Duclos18}, monolayers of epithelial cells such as Madin-Darby Canine Kidney (MDCK), human breast cancer cells (MCF-7)~\cite{Saw17}, and Human Bronchial Cells (HBC)~\cite{Blanch18}, dense cultures of amoeboid cells~\cite{Gruler99}, and neural progenitor stem cells~\cite{Kawaguchi17}.
However, unlike classic liquid crystals, these systems are `active', constantly being driven out of equilibrium by the motion of individual cells within the tissue.

There is emerging evidence that the collective dynamics of epithelial cells can be captured by theories of active liquid crystals~\cite{Prost15,Kawaguchi17,Saw17,Blanch18,Duclos18}.
For example, flows around cellular division events in epithelial cells and transition to shear flows in confined fibroblast cells have been reproduced by such a theory~\cite{Doost15,Duclos18}, and large-scale velocity fields measured in monolayers of epithelial and endothelial cells show long-range flows and patterns of localised vorticity reminiscent of the turbulent state observed in active nematic liquid crystals at high activity~\cite{doi:10.1073/pnas.0705062104, doi:10.1016/j.bpj.2010.01.030, doi:10.1038/ncomms6720,Blanch18}.
More recently, the properties of nematic topological defects, which can control death and extrusion in monolayers of MDCK cells~\cite{Saw17}, were shown to be consistent with an active nematic description.
Such a connection is highly surprising because individual epithelial cells on a substrate have a well-defined direction of movement (polarity), and cells in a monolayer are on average isotropic in shape. Therefore, it is far from obvious how the observed nematic features of the tissue can emerge from the collective dynamics.
Because most theoretical studies of epithelial cells have concentrated on polar driving~\cite{doi:10.1103/PhysRevX.6.021011, doi:10.1371/journal.pcbi.1005569} or have neglected shape deformations altogether~\cite{Gov09,Lee11,Lee11b}, much remains to be explored in order to understand how such macroscopic nematic features relate to the microscopic dynamics in cellular monolayers.

\begin{figure}[b]
    \includegraphics[width=\linewidth]{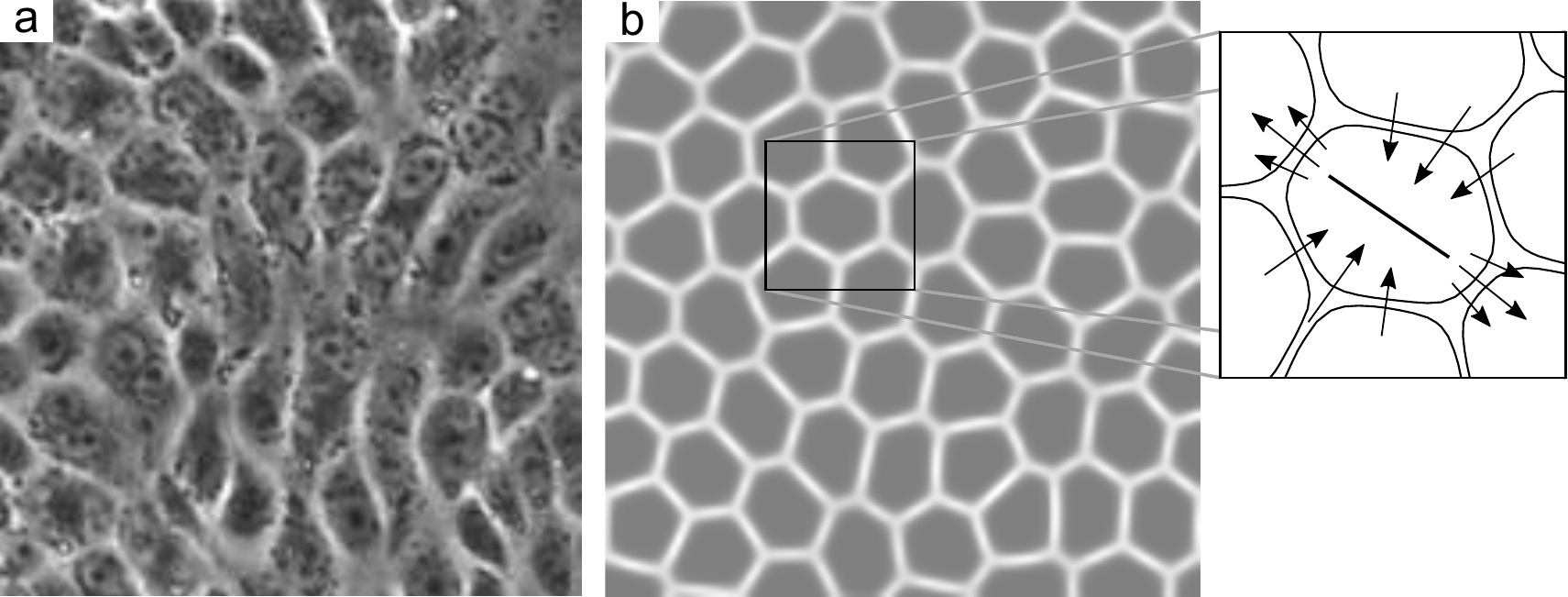}
    \caption{\label{fig:interfaces}%
        (a) A microscope image of MDCK monolayer (adapted from~\cite{Doost15}) and (b) a simulation snapshot. Interfaces between cells defined as the overlap of the phase fields $\sum_{i\neq j}\phi_i \phi_j$ (left). Contours $\phi_i=1/2$ with an illustration of the principal axis of the deformation tensor together with the resulting dipolar force density (right). The director of the dipolar force is parallel to the main deformation axis.
    }
\end{figure}

In this letter, we show that the local deformation of cells provides a suitable nematic order parameter even in systems where cells are not naturally elongated.
Using a simple microscopic model of dense two-dimensional cellular layers that captures single-cell deformations, 
we find that a minimal form of microscopic dipolar interaction between cells based on their deformations leads to spontaneous symmetry breaking and drives the system out of equilibrium.
At high activities, we also observe the proliferation of topological defects in the orientation field and the emergence of active turbulent flows as observed in monolayers of epithelial cells~\cite{Saw17,Blanch18}.
We further show that the experimentally observed patterns of flows and mechanical stresses around topological defects are accurately captured by our modeling framework.
The main goal of our study is to consider the impact of intercellular stresses that are generated due to cell-cell interactions in regulating collective behaviour in cellular tissues, as highlighted by recent experiments on epithelial cells~\cite{Ehsan18}.

We model cells as active deformable droplets in two dimensions using a phase-field model.
The phase-field approach has been widely applied to problems involving single cells (see~\cite{doi:10.1007/978-3-319-24448-8} for a review) and more recently to the description of a few migrating cells~\cite{doi:10.1103/PhysRevE.93.052405}, and the study of confluent epithelia~\cite{doi:10.1073/pnas.1219937110, doi:10.1371/journal.pcbi.1002944, Palmieri15}.
We start by describing an epithelium consisting of $N$ cells using a different phase field $\phi_i$ for each individual cell.
Values of $\phi_i = 1$ and $\phi_i = 0$ denote the interior and the exterior of a cell, respectively, and the cell boundary is defined to lie at the midpoint $\phi_i = 1/2$.
We define the dynamics of the fields $\phi_i$ as
\begin{equation}
\label{eq:dynamics}
\partial_t \phi_i + \vect v_i \cdot \nabla \phi_i = - \frac{\delta \mathcal{F}}{\delta \phi_i}, \qquad i=1, \ldots, N,
\end{equation}
where $\mathcal{F}$ is a free energy and $\vect v_i$ is the total velocity of cell $i$.

The free energy $\mathcal{F}$ defines the dynamics of the individual interfaces and is written as $\mathcal{F} = \mathcal{F}_{\text{CH}} + \mathcal{F}_{\text{area}} + \mathcal{F}_{\text{rep.}}$, where
\begin{align}
\label{eq:FE}
\begin{split}
\mathcal{F}_{\text{CH}} &= \sum_i \frac \gamma \lambda \int \dd\vect x \left\{ 4 \phi_i^2 (1-\phi_i)^2 + \lambda^2 (\nabla \phi_i)^2 \right\}, \\
\mathcal{F}_{\text{area}} &= \sum_i \mu \Big( 1 - \frac{1}{\pi R^2}\int \dd\vect x\, \phi_i^2 \Big)^2, \\
\mathcal{F}_{\text{rep.}} &= \sum_i \sum_{j \neq i} \frac{\kappa}{\lambda} \int \dd \vect x \, \phi_i^2 \phi_j^2\,.
\end{split}
\end{align}
The Cahn-Hilliard free energy $\mathcal{F}_{\text{CH}}$ stabilises the cell interface.
Our formulation is guided by simplicity but could be replaced by more realistic models of the cellular boundary~\cite{PMID:4273690, doi:10.1103/PhysRevLett.105.108104, doi:10.1103/PhysRevE.67.031908}.
The contribution $\mathcal{F}_{\text{area}}$ provides a soft constraint for the area of the individual cells around the value $\pi R^2$, where $R$ is the cell radius, such that the cells are compressible~\footnote{Even though epithelial cells are rather incompressible in 3D, they are effectively compressible in 2D because they can stretch in the direction normal to the substrate plane.}
Finally, the repulsion term $\mathcal{F}_{\text{rep.}}$ penalises regions where two cells overlap.
Normalisation has been chosen such that the width of the interfaces at equilibrium is $\lambda$ and such that the properties of the cells are roughly preserved when $\lambda$ is rescaled (see SI).
The parameters $\gamma$, $\mu$, and $\kappa$ set the relaxation time scale of shape deformations, area changes, and repulsive forces, respectively (see SI for the parameter values).

This formulation allows the cellular interfaces to be resolved, and intracellular forces to be defined at the microscopic level (Fig.~\ref{fig:interfaces}b).
There are many physical forces of importance at the cellular level~\cite{doi:10.1186/s12915-015-0150-4, doi:10.1038/ncb3564}, and force transmission between cells has been shown to contribute to collective phenomena such as collective durotaxis~\cite{doi:10.1126/science.aaf7119} or coordination during morphogenesis~\cite{Guillot13, Etournay2015} and wound healing~\cite{doi:10.1126/science.1221071, doi:10.1038/nphys3040}.
We concentrate here on a simplified description and consider only substrate friction and forces generated at the cellular interfaces, leading to the force balance equation:
\begin{equation}
\label{eq:force balance}
\xi \vect v_i = \vect F^{\text{int.}}_i,
\end{equation}
where $\xi$ is a substrate friction coefficient and $\vect F^{\text{int.}}_i$ is the total force acting on the interface of cell $i$. In an analogy with continuum theories, we define these microscopic interface forces in terms of a macroscopic tissue stress tensor $\mathrm \sigma_{\text{tissue}}$ as
\begin{equation}
\label{eq:Ftot}
\vect F^{\text{int.}}_i = \int \dd \vect x\, \phi_i \; \nabla \cdot \mathrm \sigma_{\text{tissue}} = - \int \dd \vect x\, \mathrm \sigma_{\text{tissue}} \cdot \nabla \phi_i.
\end{equation}
The first expression is the integral of the local force $\nabla\cdot \mathrm \sigma_{\text{tissue}}$ weighted by the phase field $\phi_i$, while the second is the integral of the force exerted by the stress tensor on the vector $-\nabla \phi_i$ normal to the interface and pointing outwards.
Equation~\eqref{eq:Ftot} effectively bridges scales between local forces at the level of the individual cells and properties of the whole tissue.


Equations~(\ref{eq:dynamics})--(\ref{eq:Ftot}) define a generic model of two-dimensional epithelial monolayers that only requires an appropriate definition of the stress tensor as input.
Following our analogy with continuum theories, we introduce the usual separation into passive and active stresses by writing
\begin{equation} \label{eq:zeta}
\matr \mathrm \sigma_{\text{tissue}} = - P \mathbb I - \zeta \matr Q,
\end{equation}
where the fields $P$ and $\matr Q$ are the tissue pressure and tissue nematic tensor to be defined below.
As pointed out in~\cite{Palmieri15}, there is in fact much freedom in defining $P$ from the total free energy $\mathcal{F}$.
Here we use the thermodynamically consistent definition
\begin{equation}
\label{eq:pressure}
P = \sum_{i}\left( \frac{\delta \mathcal{F}_{\text{rep.}}}{\delta \phi_i} - \frac{\delta \mathcal{F}_{\text{CH}}}{\delta \phi_i} - \frac{\delta \mathcal{F}_{\text{area}}}{\delta \phi_i} \right),
\end{equation}
which includes contributions from compression and surface tension terms (see SI for details).
%
\begin{figure}[t]
    \raisebox{1.8in}{a}\hspace*{-.3em}
    \hspace{.75cm}\includegraphics[width=3in]{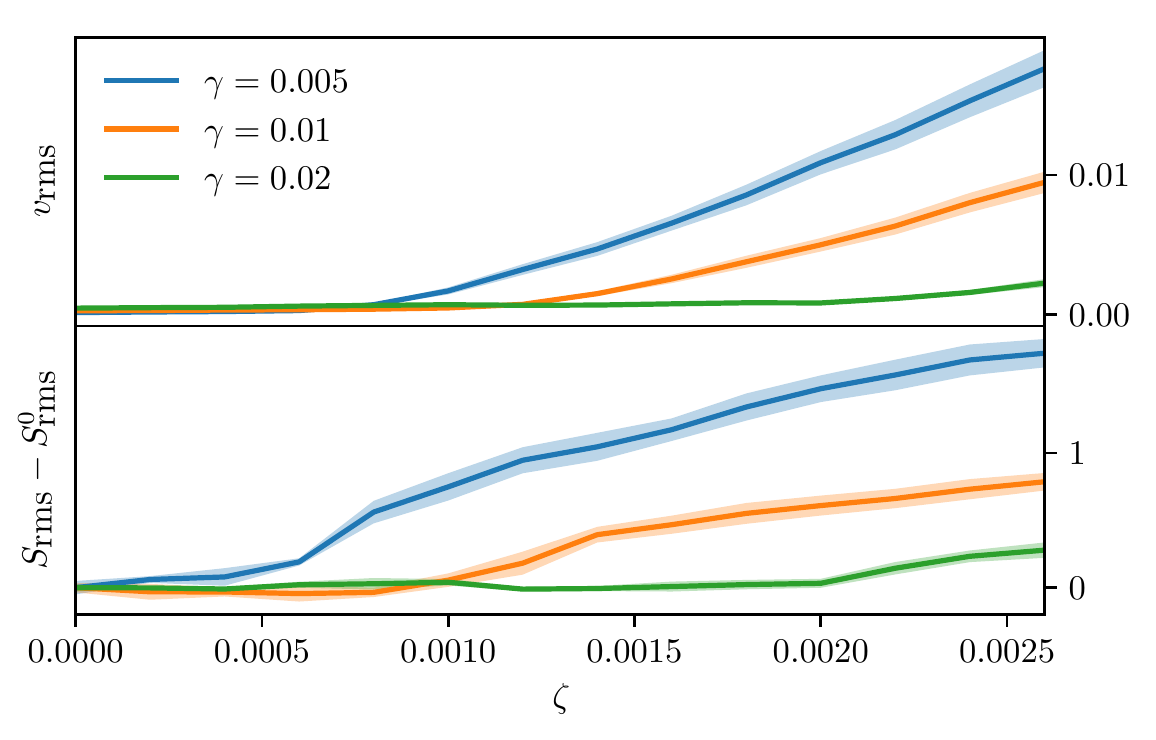}\\
    \raisebox{1.2in}{b}\hspace*{-.2em}
    \includegraphics[width=3.1in]{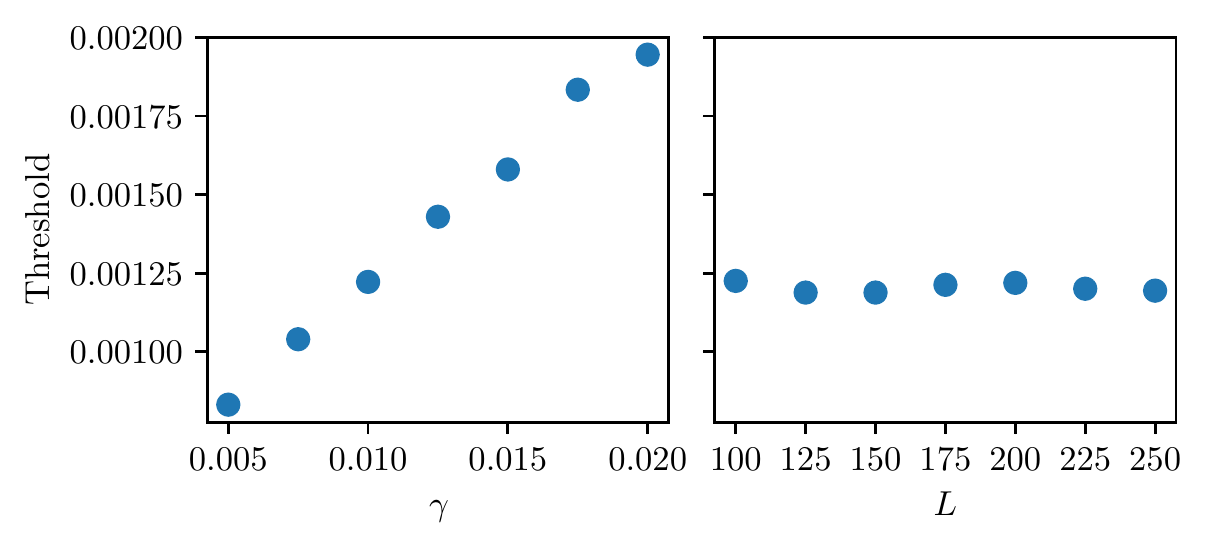}\hspace*{4cm}
    \caption{\label{fig:transition}%
        Transition to flows in an extensile system of 240 cells. 
        (a) The root-mean-square velocity $v_{\textrm{rms}}$ and nematic order $S_{\textrm{rms}}$ develop non-zero values with increasing activity $\zeta$, where the mean value is taken over each individual cell.
        Note that we subtracted the zero-activity rms order $S^0_{\textrm{rms}} = S_{\textrm{rms}}|_{\zeta=0}$ and that we have smoothed both fields using a sliding window of size $3R \times 3R$.
        Mean$\pm$std from 5 simulations.
        (b) Dependence of the location of the activity threshold on the elasticity $\gamma$ and domain size $L$.
        The threshold is defined as the first value of activity $\zeta$ for which the $v_{\text{rms}} > 10^{-7}$ after 3000 simulation steps, after $v_{\text{rms}}$ has been averaged over 4 simulations.
    }
\end{figure}
Key to our results is the definition of the tissue nematic tensor in Eq.~\eqref{eq:zeta}, which is based on the cell deformation:
\begin{equation}
\label{eq:Q}
\matr Q = \sum_i \phi_i \matr S_i,
\end{equation}
where $\matr S_i$ is the deformation tensor of cell $i$ defined as the traceless part of $- \int\dd\vect x\, (\nabla \phi_i)^\transp \nabla \phi_i$.
Its eigenvalues and eigenvectors measure the strength and orientation of the main deformation axes of cell $i$ (see SI for details).
Multiplying each deformation tensor by the corresponding phase field ensures that $\matr Q$ is a field defined at each point in space.

We now turn to the main focus of this paper and show that introducing a local active term proportional to the deformation of single cells is able to drive the system out of equilibrium and to capture the active nematic phenomenology of dense cellular monolayers. From Eq.~\eqref{eq:Ftot}, one can see that the active term, $\zeta\matr Q$, for a given nematic tensor $\matr Q$ can be interpreted as a dipolar force density distributed along the cells' interfaces while $P$ is a simple elastic repulsion force.
As a result, each cell pushes or pulls its neighbours depending on the direction of their contact area with respect to the stress tensor (see Fig.~\ref{fig:interfaces} for an illustration).
Note however that this is an effect of the cellular interactions alone, such that single, isolated, cells do not deform at non-vanishing activity.

\begin{figure}[h!]
    \includegraphics[width=3in]{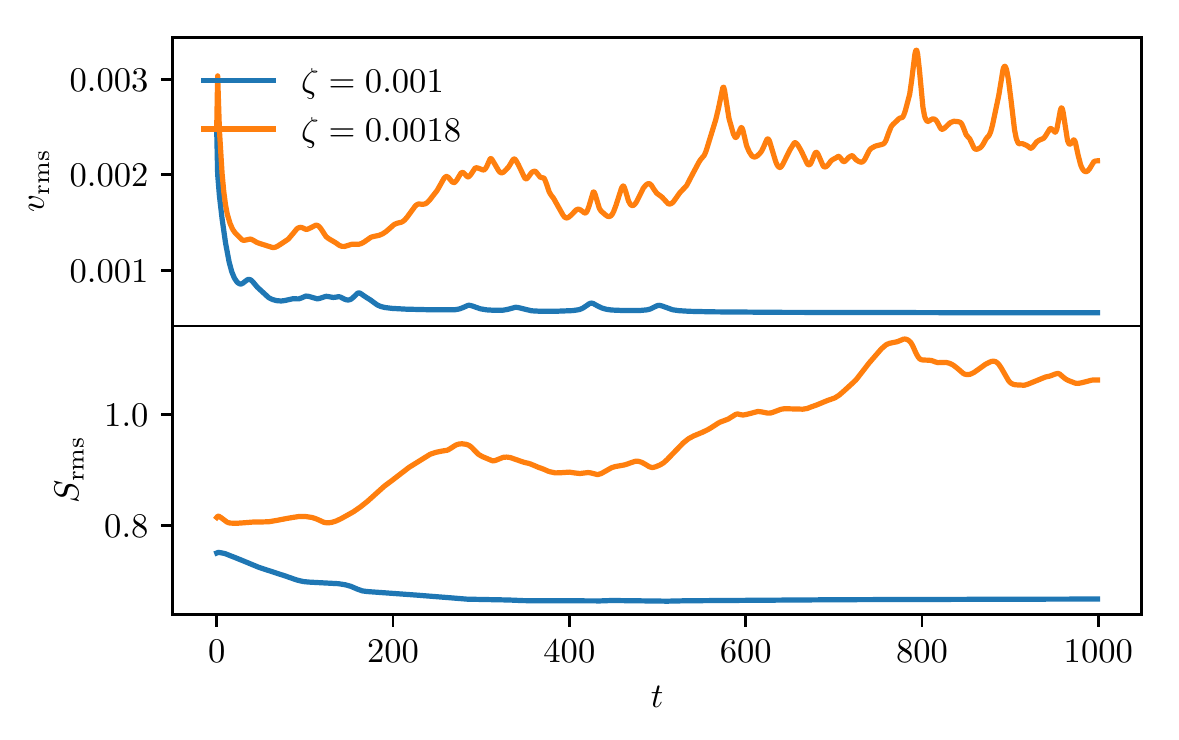}
    \caption{%
        \label{fig:time dep}
        Temporal evolution of the root-mean-square velocity and order above (orange) and below (blue) threshold.
        The initial relaxation from the base configuration is apparent in both cases.
        The system then relaxes completely to its equilibrium when the activity is below threshold, while it is able to sustain large flows and shows non vanishing increase its order above threshold.
        Note that even at equilibrium, there is a residual order from the deformation of cells at close packing.}
\end{figure}

This simple definition of the tissue nematic tensor in terms of the local deformation is able to create large-scale flows for high enough activity strengths and leads to the spontaneous creation of defects in the nematic field (see Suppl. Movie~1).
Defining the tissue velocity as $\vect v = \sum_i \phi_i \vect v_i$, we see that the root-mean-square velocity $v^2_{\textrm{rms}} = \langle \vect v ^2\rangle$ and  nematic order $S^2_{\textrm{rms}} = \langle \det \matr S^2 \rangle = \langle S_{11}^2 + S_{12}^2 \rangle$ develop non-zero values as $\zeta$ is increased (Fig.~\ref{fig:transition}(a) and Suppl. Fig.~\ref{fig:time dep}), indicating that our model shows an activity-driven transition to non-zero nematic order and flows. 
Because of the nematic nature of the interactions, the total force is approximately zero at the tissue level and the system does not develop any system-wide net velocity under periodic boundary conditions. In particular this means the transition to collective movement here is different from the flocking phase transitions observed in Vicsek-type active systems with polar driving.

\begin{figure}[h!]
    \includegraphics[width=3in]{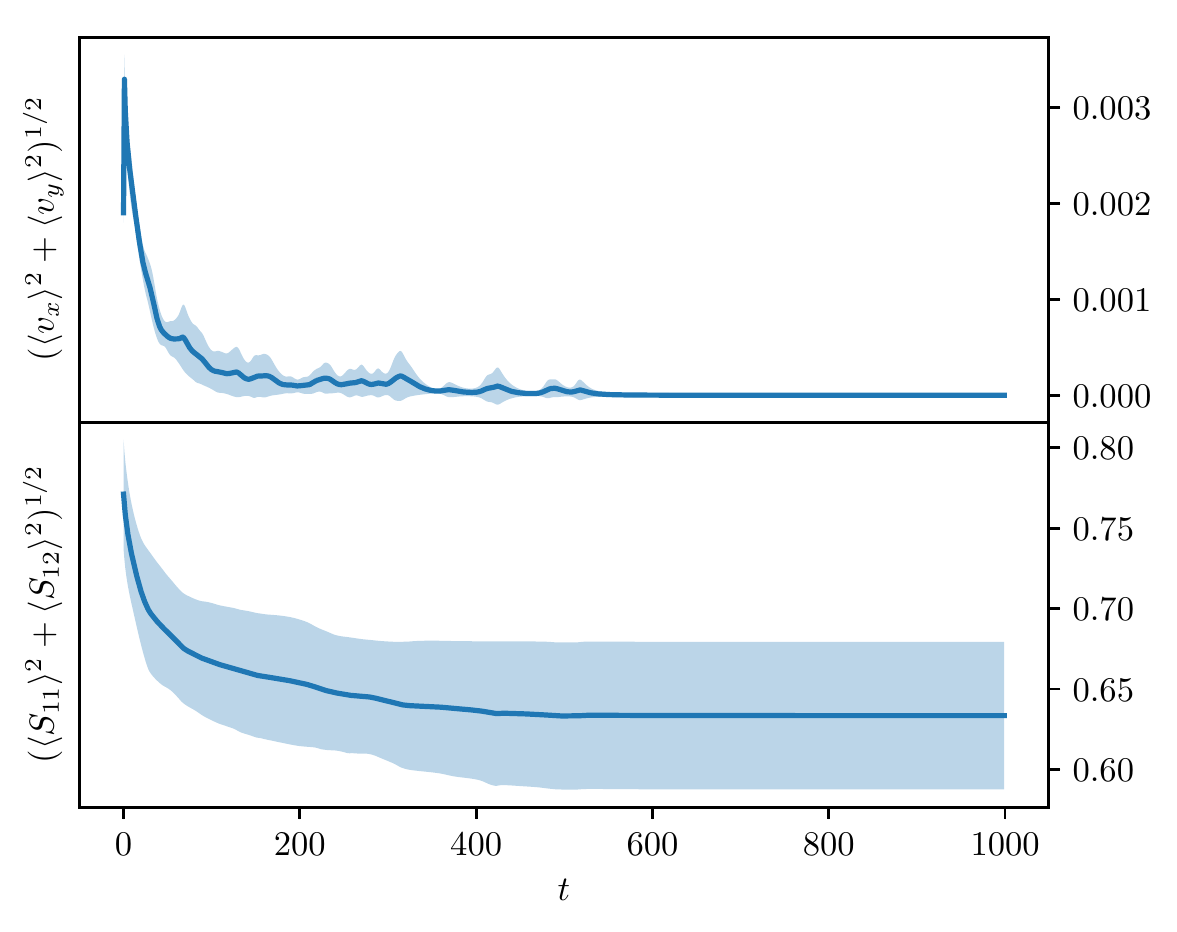}
    \caption{
        \label{fig:time dep2}
        Evolution of the root-mean square velocity and order in a system with extensile activity $\zeta = -0.0018$ shows that the model is stable in this case.
            Mean$\pm$std from 5 simulations.
    }
\end{figure}

The transition to flow shows a clear activity threshold which depends in a well-defined manner on the elasticity $\gamma$ (Fig.~\ref{fig:transition}(b)), but is independent of the domain size $L$ (Fig.~\ref{fig:transition}(c)). This is in contrast to continuum models of active nematics, which explicitly include the velocity as a hydrodynamic variable and where a hydrodynamic instability initiates flows at an activity threshold that tends to zero as $L \rightarrow \infty$~\cite{doi:10.1103/physrevlett.89.058101}.
This suggests a different origin for the transition and we conjecture that it is driven by the interaction forces at the cell-cell interfaces that amplify the deformation for extensile activity ($\zeta>0$) in our model.
This hypothesis is strengthened by the fact that the system is stable for contractile activity ($\zeta<0$) for which the interactions at the interfaces tend to restore the equilibrium shape (Suppl. Fig.~\ref{fig:time dep2}).
A finite value of the threshold then appears because cells need to push (pull) strongly enough to cause sufficient deformation of their neighbours. Using the balance of active stresses with the pressure contribution due to the elastic energetic cost of cell deformation, the activity threshold is found to linearly depend on the elasticity, $\zeta_{\text{cr}}\sim\gamma$, consistent with the simulation results (Fig.~\ref{fig:transition}(b)), see SI for more details.

These differences in the origin of the spontaneous flow generation have clear observable consequences for the dynamics of the system.
In particular, we do not observe the formation of highly distorted lines in the nematic field --- `walls' --- that are a typical consequence of the hydrodynamic instability~\cite{Giomi13,Thampi14}.
This in turn affects the mechanism by which nematic defects are spontaneously created.
While in continuum active nematics defect proliferation is predominantly mediated by the unzipping of walls by pairs of oppositely charged defects~\cite{Giomi13}, the emergence of defect pairs occurs spontaneously at random positions in our cell-based model.
Interestingly this 
resembles much more closely the experimental observations of the defect proliferation in MDCK cells where it has been puzzling that no walls are apparent~\cite{Saw17}.
\begin{figure}[t]
    \raisebox{1.8in}{a}\hspace*{-.3em}
    \includegraphics[width=3in]{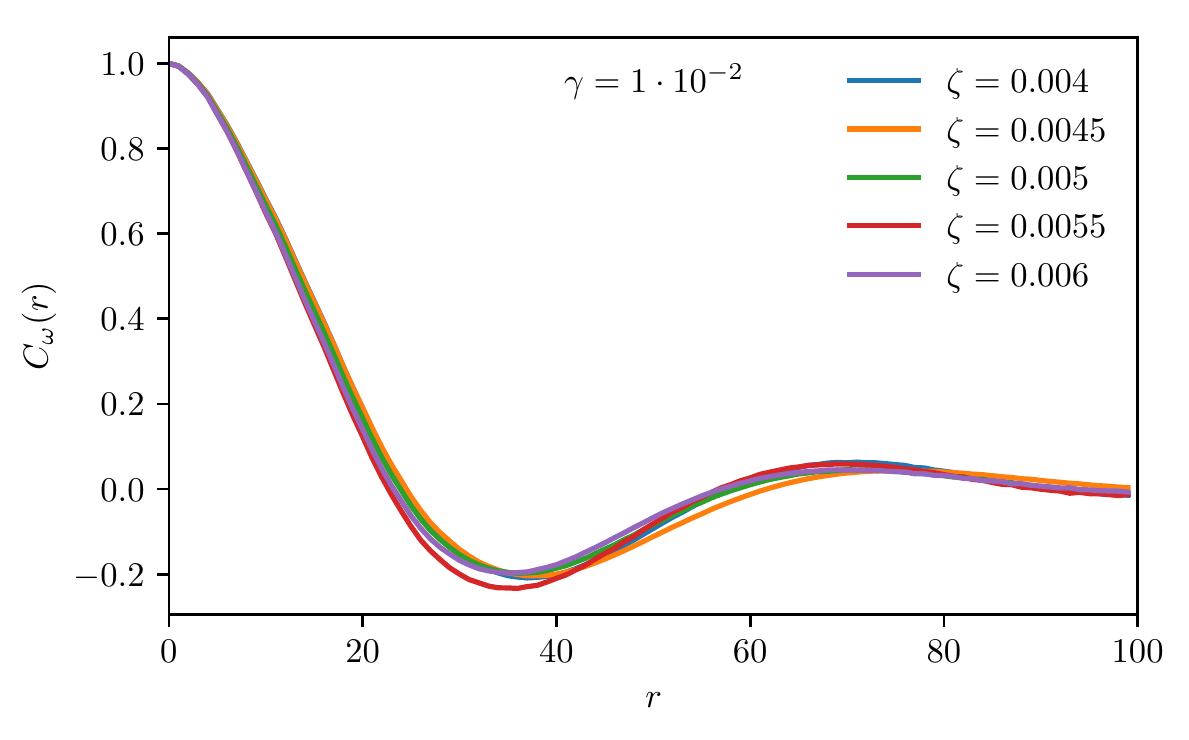} \\
    \raisebox{1in}{b}\hspace*{-.3em}
    \includegraphics[width=1.6in]{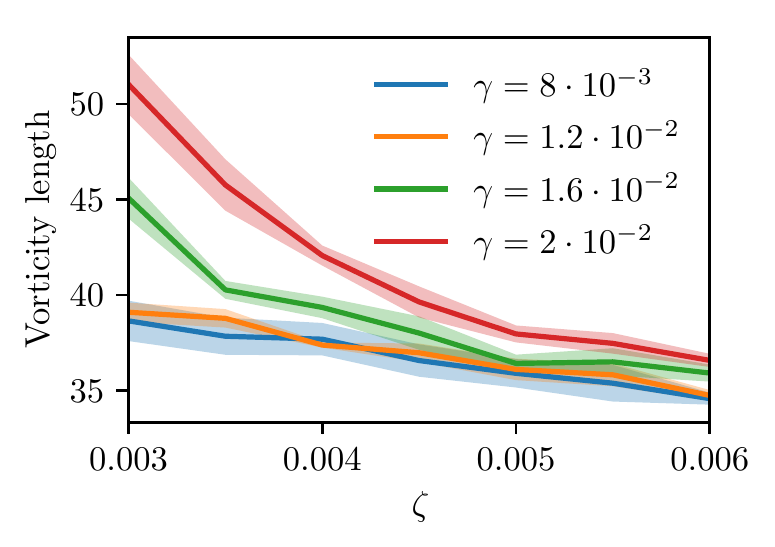}
    \raisebox{1in}{c}\hspace*{-.3em}
    \includegraphics[width=1.6in]{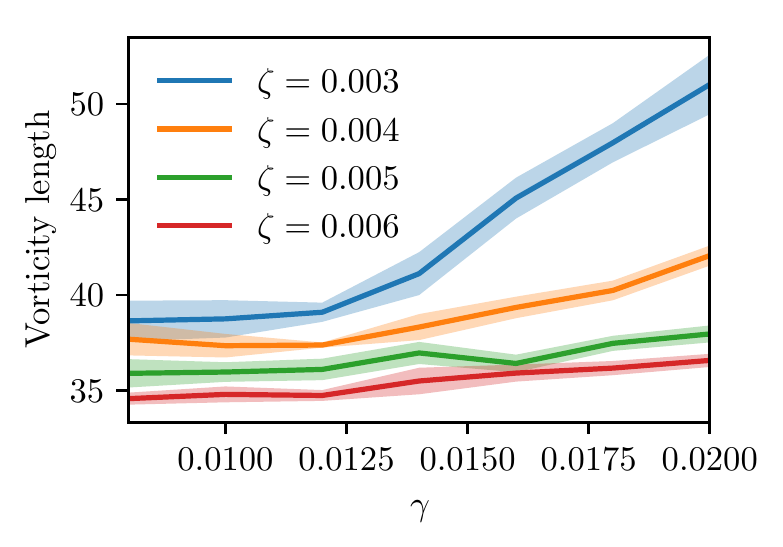}\\
    \caption{\label{fig:turbulence}%
        Properties of flows during active turbulence for an extensile system.
        (a) Spatial autocorrelation function $C_\omega(R) = \langle \omega(R) \omega(0) \rangle/\langle \omega(0)^2\rangle$ of the vorticity for different values of $\zeta$.
        The vorticity field has been smoothed using a sliding window of size $3R \times 3R$.
        (b--c) Dependence of the vorticity length, defined as the location of the minimum of the vorticity autocorrelation function, as a function of activity $\zeta$ and elasticity $\gamma$.
        Mean$\pm$std from 5 simulations.
    }
\end{figure}

\begin{figure}[h!]
    \includegraphics[width=3in]{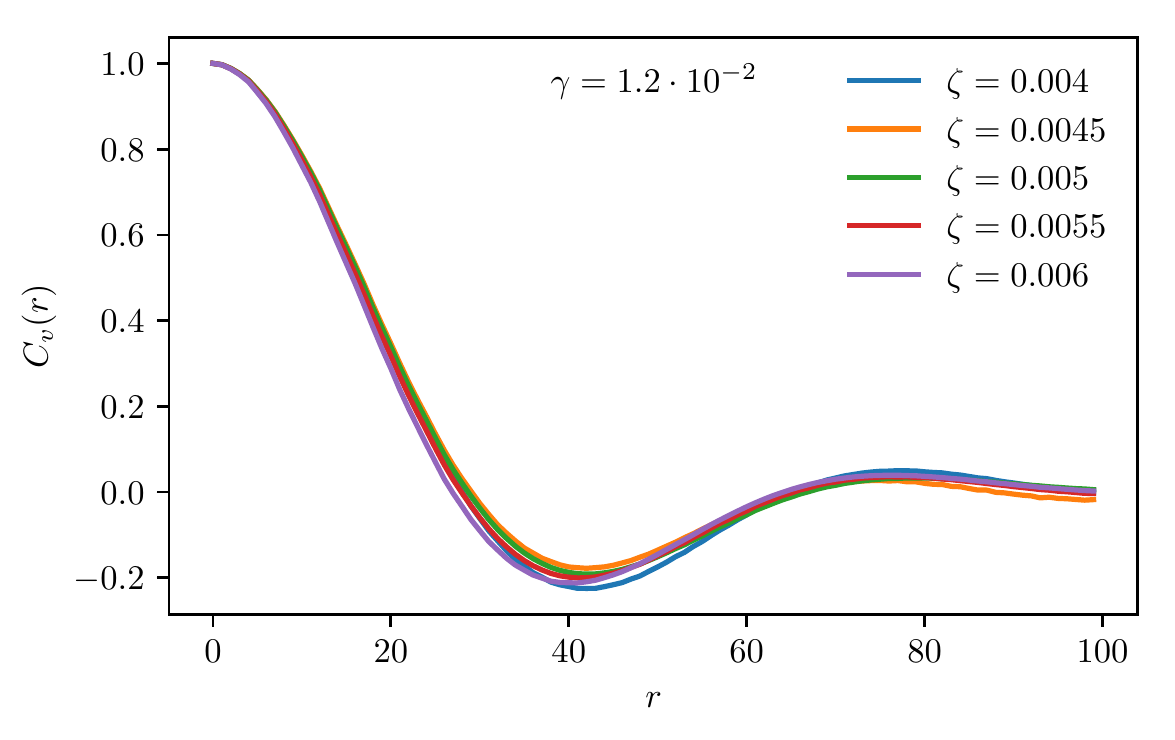}\\
    \includegraphics[width=3in]{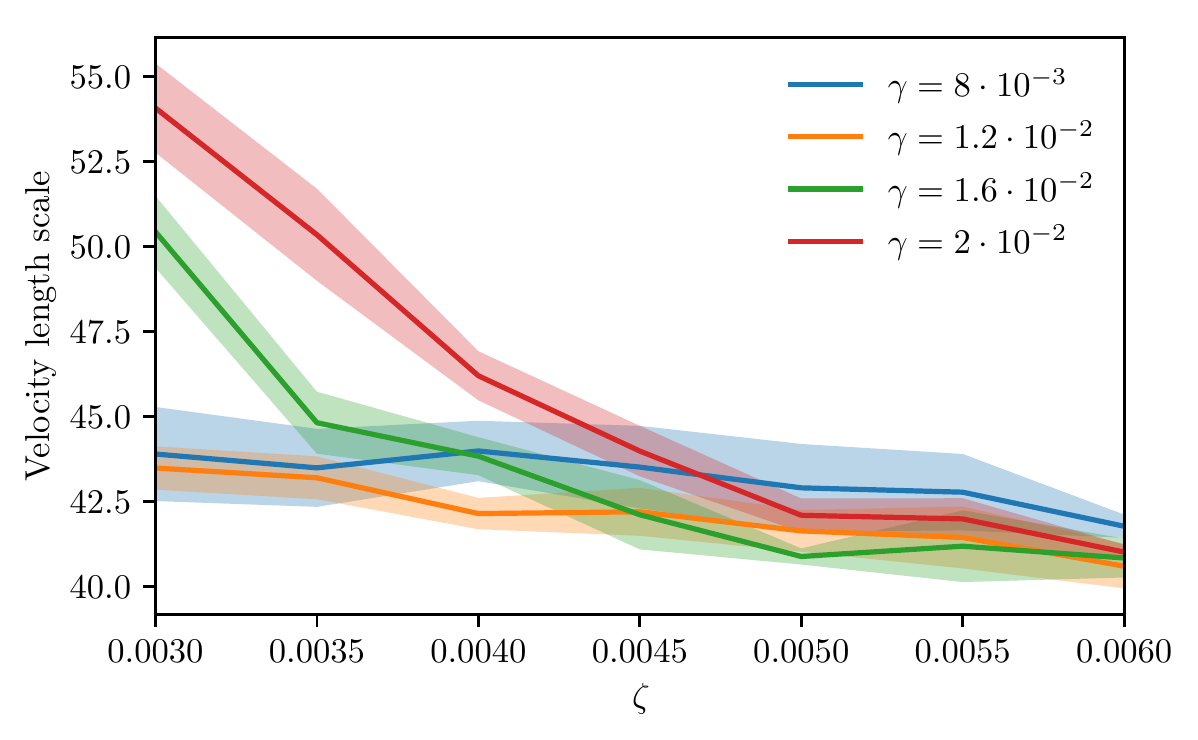}
    \includegraphics[width=3in]{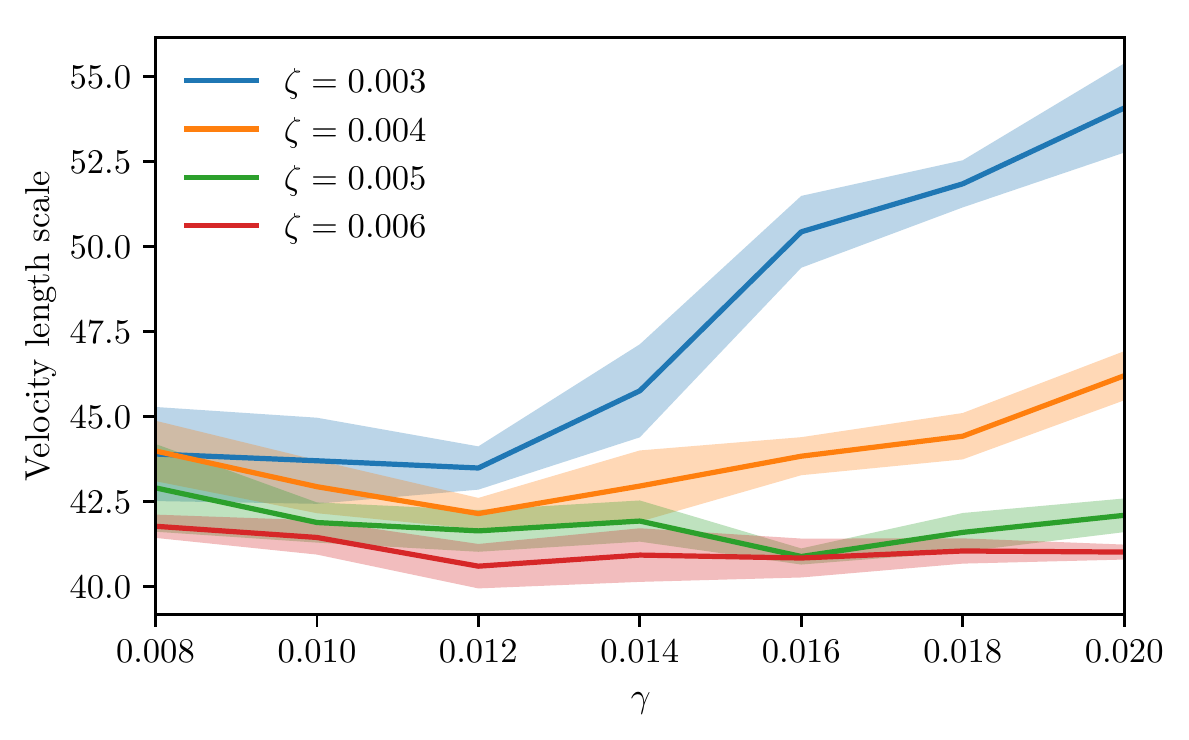}
    \caption{\label{fig:vel-length}%
        Spatial autocorrelation function $C_v(R) = \langle \vect v(R) \vect v(0) \rangle/\langle \vect v(0)^2\rangle$ of the velocity for different values of $\zeta$ (top).
        The velocity field has been smoothed using a sliding window of size $3R \times 3R$.
        Dependence of the velocity length scale defined as the location of the minimum of the velocity autocorrelation function as a function of $\zeta$ (left) and $\gamma$ (right).
        Mean$\pm$std from 5 simulations.
    }
\end{figure}

\begin{figure}[h!]
    \includegraphics[width=3in]{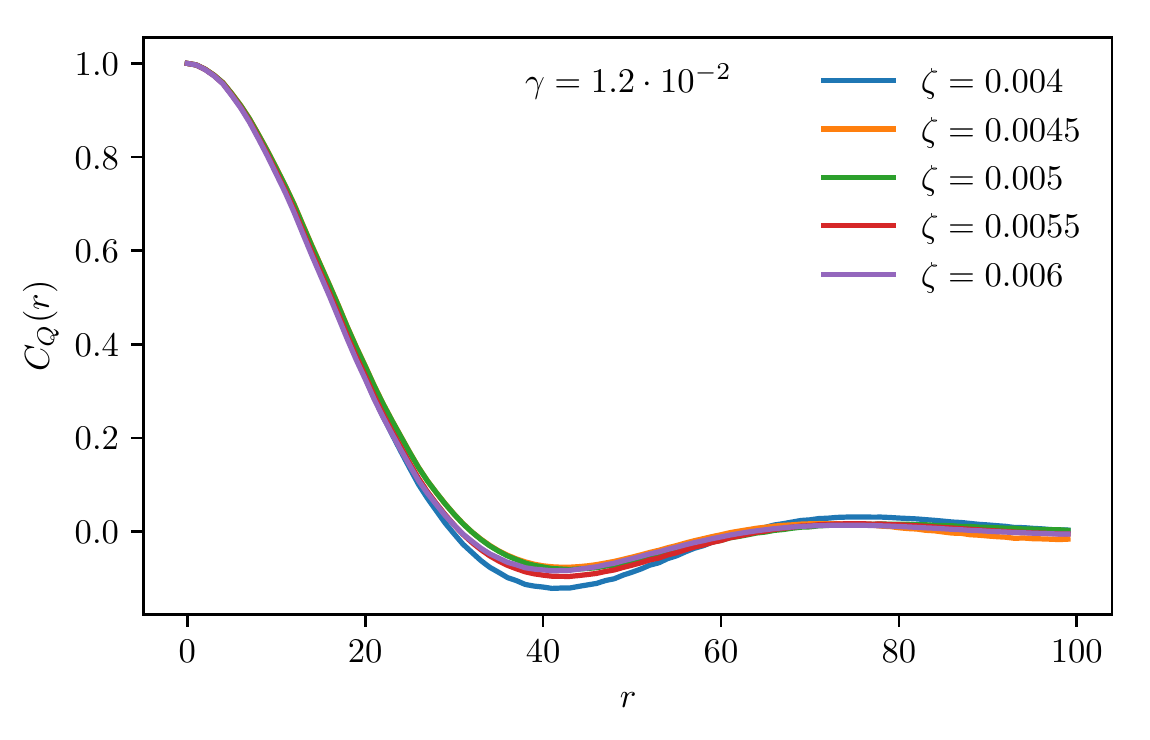}\\
    \includegraphics[width=3in]{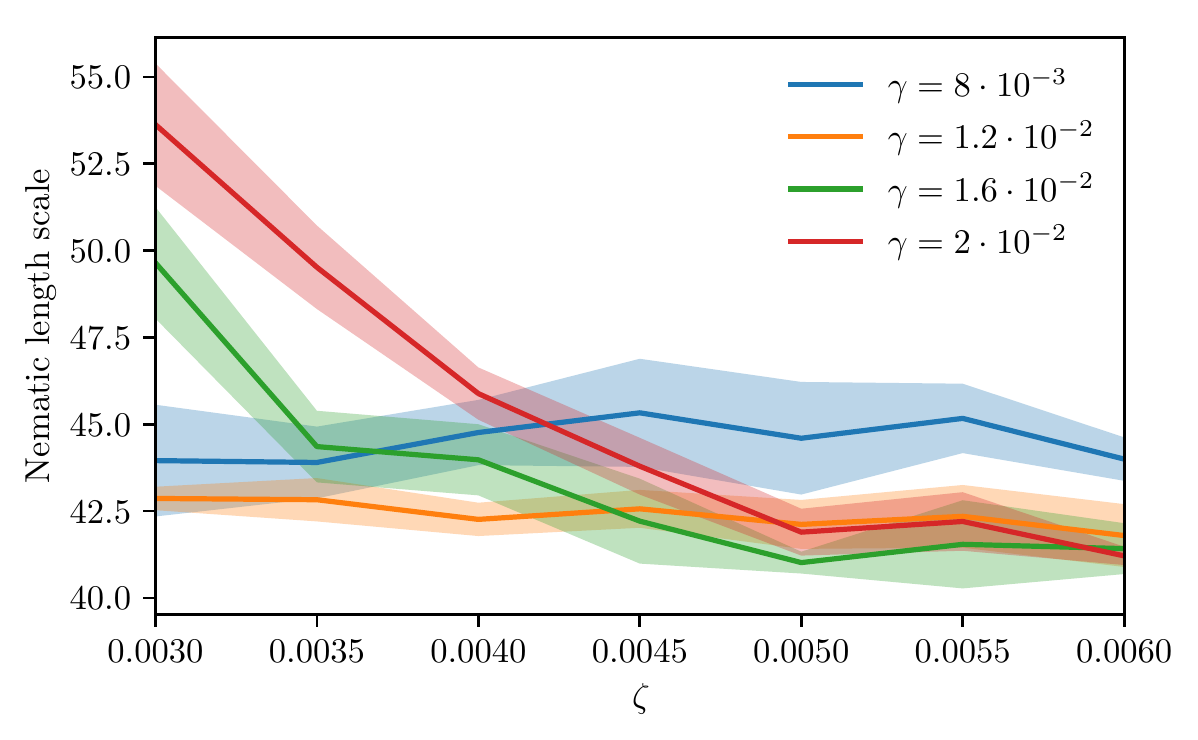}
    \includegraphics[width=3in]{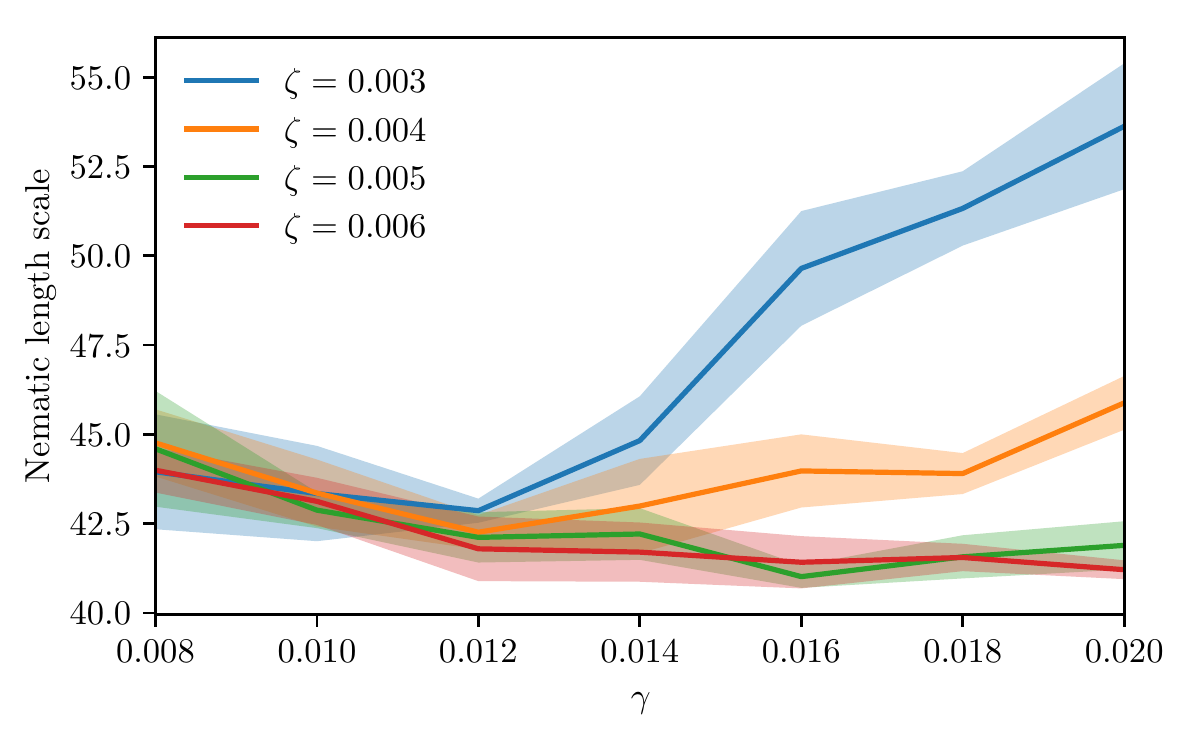}
    \caption{\label{fig:nem-length}%
        Spatial autocorrelation function $C_Q(R) = \langle Q_{11}(R) Q_{11}(0) + Q_{12}(R) Q_{12}(0) \rangle/\langle Q^2_{11}(0) + Q^2_{12}(0)\rangle$ of the nematic tensor for different values of $\zeta$ (top).
        The nematic field has been smoothed using a sliding window of size $3R \times 3R$.
        Dependence of the nematic length scale defined as the location of the minimum of the nematic autocorrelation function as a function of $\zeta$ (left) and $\gamma$ (right).
        Mean$\pm$std from 5 simulations.
    }
\end{figure}

\begin{figure}[h!]
    \raisebox{1.8in}{a}\hspace*{-.3em}
    \includegraphics[width=3in]{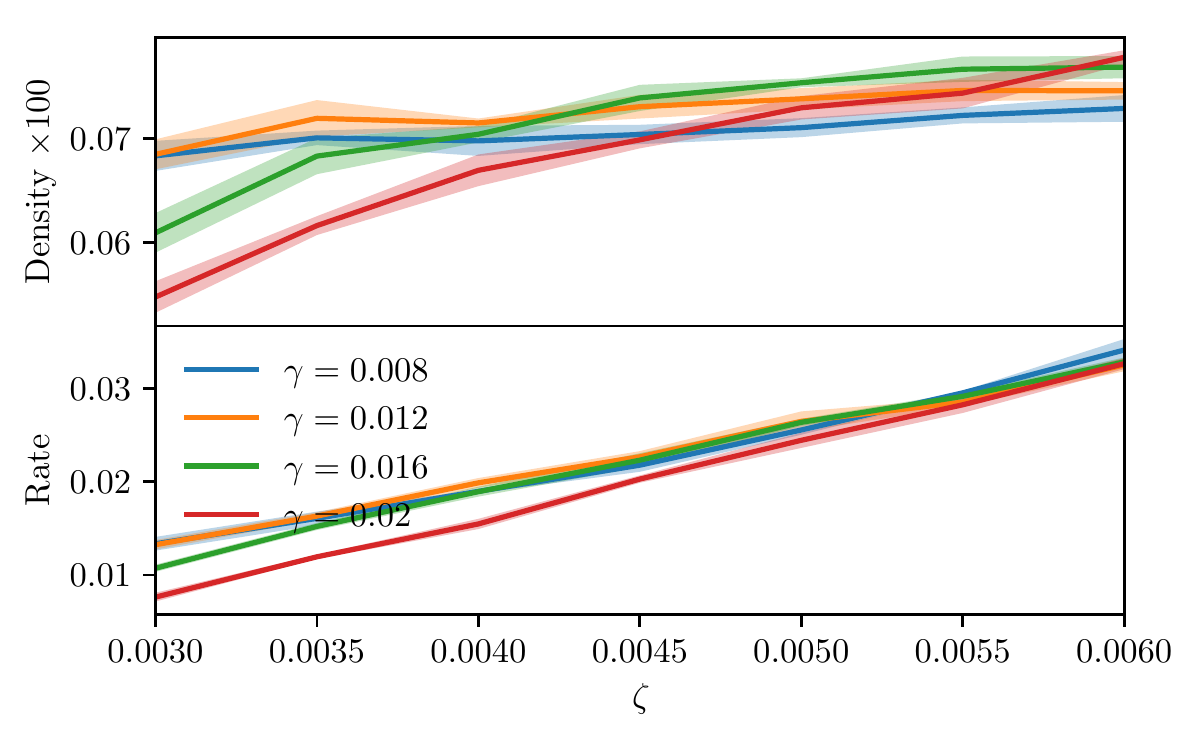}
    \raisebox{1.8in}{b}\hspace*{-.3em}
    \includegraphics[width=3in]{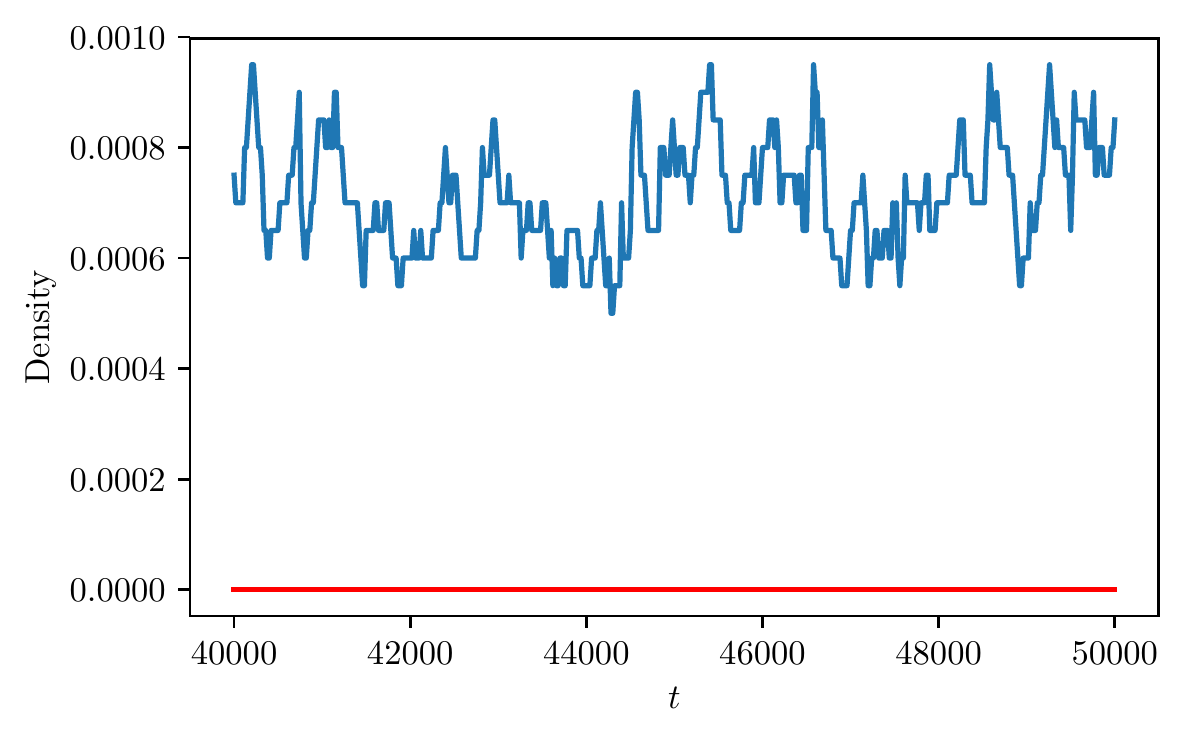}
    \caption{\label{fig:moar defects}
        (a) Defect density and rate of defect creation for different values of the activity $\zeta$.
            The rate shows a marked increase with increasing activity $\zeta$ in agreement with continuum theories of active liquid crystals.
            Mean$\pm$std from $5$ simulations.
            (b) The temporal evolution of the total density of defects (positive plus negative defects, blue) and total density of charge (positive minus negative defects, red) during turbulence shows that even though the total number of defects changes with time, they are always created or destroyed in pairs such that the total charge remains zero in the system.
    }
\end{figure}

Spontaneous defect pair creation leads to the emergence of a state resembling the active turbulence observed in continuum theories of active nematic liquid crystals~\cite{Giomi13} (see Suppl. Movie~2).
The vorticity autocorrelation function $C_\omega(r)=\langle\omega(r,t)\cdot\omega(0,t)\rangle/\langle\omega(0,t)^{2}\rangle$ for different values of the activity $\zeta$ shows a well-defined  length-scale determined by the minimum of $C_\omega(r)$ (Fig.~\ref{fig:turbulence}(a)).
Autocorrelation functions for the velocity and the nematic field show similar behaviour (see Suppl. Figs.~\ref{fig:vel-length} and~\ref{fig:nem-length}).
This indicates that force transmission mediated by cell-cell contacts leads to long-range flows at macroscopic scales.
Increasing activity $\zeta$ leads to smaller vortices, while increasing elasticity $\gamma$ results in larger vortices (Fig.~\ref{fig:turbulence}(b) and (c)). In addition, consistent with continuum hydrodynamic models~\cite{Giomi15}, defects are always created or destroyed in pairs and defect density and defect creation rate both increase with increasing activity $\zeta$ (Suppl. Fig.~\ref{fig:moar defects}).

\begin{figure}[t]
    \raisebox{1.5in}{a}\hspace*{-.3em}
    \includegraphics[width=2.5in]{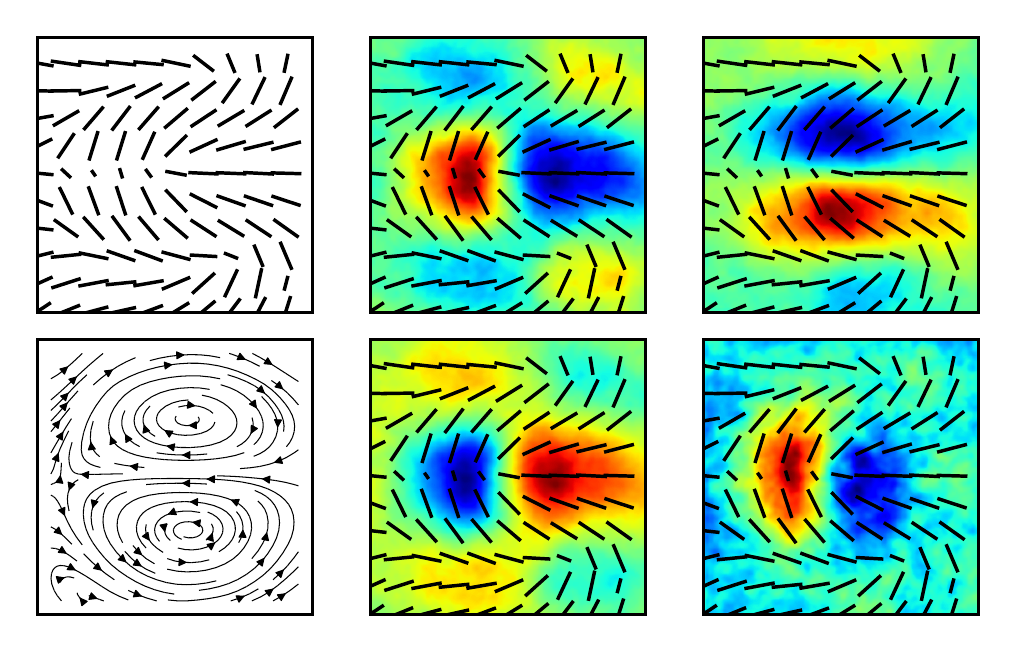} \\
    \raisebox{1.5in}{b}\hspace*{-.3em}
    \includegraphics[width=2.5in]{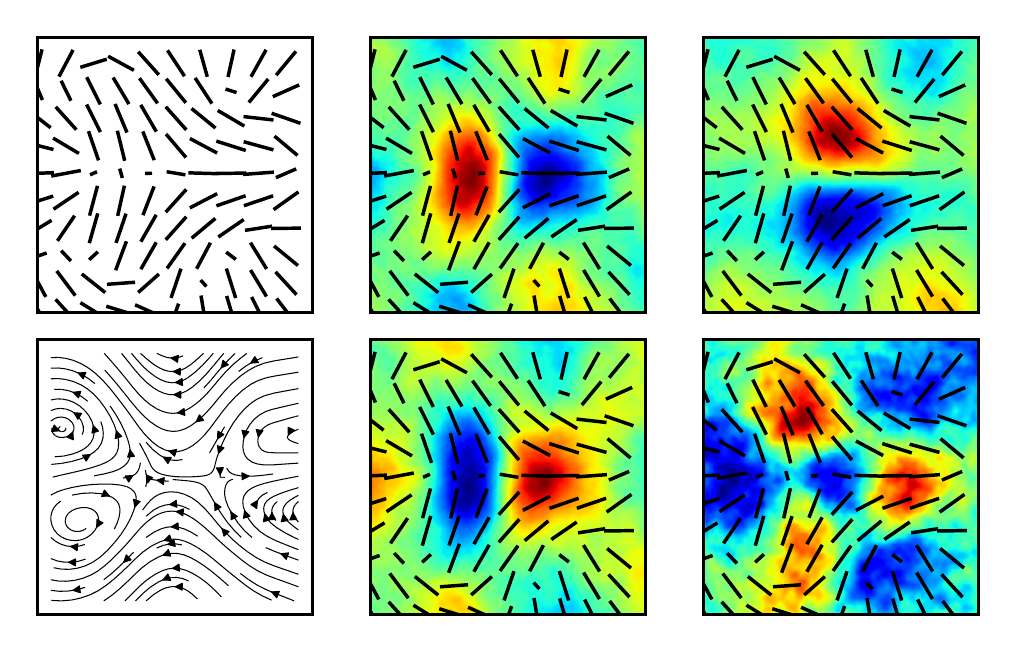}
    \caption{\label{fig:defect}%
        Average properties of $+1/2$ (a) and $-1/2$ (b) defects in an extensile system with $\zeta=5 \cdot 10^{-3}$ and $\gamma=1.4\cdot 10^{-2}$: nematic field (top left), flow field (bottom left), $\sigma_{xx}$ (top center), $\sigma_{yy}$ (bottom center), $\sigma_{xy}$ (top right), pressure (bottom right).
        Colors are normalized such that the maximum value is red and the minimum value is blue.
        Average over 5 simulations and each box shows a domain of size $100 \times 100$ and corresponds to approximately 60 cells.
    }
\end{figure}

Finally, we analysed the properties of flows and stresses around defects, which are crucial in determining the dynamics of active turbulence~\cite{Giomi13}.
Figure~\ref{fig:defect} shows the isotropic stress patterns and flow fields around $\pm1/2$ defects obtained from the cell-based model. These agree very well with the analytical prediction of flow fields around isolated defects~\cite{Giomi12, Giomi14}
, as well as with recent experimental measurements of defects flow fields and isotropic stresses in epithelial monolayers~\cite{Saw17}.

Together, our results demonstrate that a cell-based model which accounts for cell deformability and force transmission at cell-cell contacts can serve as a minimal and generic model to explain the active liquid crystal properties found recently in epithelial monolayers.
The model reproduces the phenomenology of active liquid crystals, together with mechanical stress and flow patterns consistent with experimental measurements.
It explains that a bootstrap feedback between shape deformations and intercellular driving allows cells of isotropic shape, such as MDCK cells or non-aggressive human breast cancer cells (MCF-7), to drive an instability to spontaneous flow and to create topological defects.
Even though our model is based on a force-balance formulation, in the macroscopic limit it shows clear hydrodynamic behaviour due to the long-range interactions mediated by cell-cell contacts.

Our minimal model leads to a number of testable predictions that could challenge current understanding in tissue mechanics. Even though individual cells are internally made of contractile material~\cite{Pomp18}, there is now clear evidence that their macroscopic behaviour can show properties of extensile nematic theories~\cite{Saw17,Blanch18}, but the underlying mechanism remains controversial.
Here, we predict that such a coarse-grained extensility can arise from interactions between the cells. 
High resolution measurements of the forces acting on individual cells within a confluent monolayer can test this theory and further shed light on the force transmission mechanism. 
Furthermore, the existence of an activity threshold can be examined in experiments by introducing incremental dosages of drugs that impair the cell motility and molecular perturbations that affect cell deformability.

\section*{Acknowledgements}
R.M. was supported by grant P2EZP2\_165261 of the Swiss National Science Foundation.
A.D. was supported by a Royal Commission for the Exhibition of 1851 Research Fellowship.

\addcontentsline{toc}{chapter}{Bibliography}
\bibliography{bibliography} 

\newpage
\thispagestyle{empty}
\renewcommand{\figurename}{SUPP. FIG.}

\maketitle

\onecolumngrid

\section{Supplementary information}

\subsection{Scaling with respect to the interface width}
All the quantities appearing in the main text have been rescaled with respect to $\lambda$ in order to ensure that the model is approximately invariant under changes of the interface width.
The equilibrium shape of the interfaces is obtained by minimising the Cahn-Hilliard free energy $F_{\text{CH}}$.
Considering only phase fields invariant under translation in the $y$ direction, it is easy to see that $\phi_\pm (x) = \left(1 \pm \tanh\left( x / \lambda\right) \right)/2$ satisfy the corresponding Euler-Lagrange equation.
These two solutions correspond to equilibrium interfaces of width $\lambda$ and of infinite length along the $y$ direction with boundary conditions $\phi_\pm(\pm \infty) = 1$ and $\phi_\pm(\mp \infty) = 0$.
We can then obtain the scaling with respect to $\lambda$ of the following expressions:
\begin{align}
\label{eq:scalings}
\begin{split}
\int \dd x\, \phi_\pm(x) \phi_\mp(x) &= \lambda/2, \\
\int \dd x\, \phi^2_\pm(x) \phi^2_\mp(x) &= \lambda/12, \\
\int \dd x\, \phi_\pm(x) \partial_x \phi_\mp(x) &= 1/2, \\
\int \dd x\, \phi^2_\pm(x) \partial_x \phi_\mp(x) &= 1/3,
\end{split}
\end{align}
and rescale all quantities in the model accordingly.
Note that it does not mean that the model is completely invariant under rescaling of $\lambda$ because the repulsion free-energy term $F_{\text{rep.}}$ changes slightly the shape of the interface and because the dynamics is not necessarily close to equilibrium for non-zero activity.
In general, repulsion forces tend to reduce the interface size such that the actual overlap between cells is measured by an effective interface width $\lambda^* < \lambda$ that depends on the repulsion parameter $\kappa$ as well as the local density of cells.

\subsection{Definition of tissue pressure}
As noted in \cite{Palmieri15}, the expression $\frac{\delta F}{\delta \phi_i} \nabla \phi_i$ can be interpreted as a local force density on the interface of cell $i$.
Note that the direction of the force is given by $\nabla \phi_i$ and hence is perpendicular to the cell interface and pointing inwards.
Decomposing the different contributions gives
\begin{align*}
\frac{\delta F_{\text{CH}}}{\delta \phi_i} &= \frac{8 \gamma} \lambda \phi_i (1-\phi_i)(1-2\phi_i) - 2 \gamma \lambda \Delta \phi_i, \\
\frac{\delta F_{\text{area}}}{\delta \phi_i} &= - \frac{4 \mu}{\pi R^2} \phi_i \left(1 - \frac 1 {\pi R^2} \int \dd \vect x\, \phi_i^2 \right), \\
\frac{\delta F_{\text{rep.}}}{\delta \phi_i} &= \frac{2 \kappa} \lambda \sum_{k\neq i} \phi_k^2 \phi_i.
\end{align*}
The first two contributions arise from the surface tension of the interface and compression of the cells, respectively.
They describe the forces created by the internal mechanical properties of a cell $i$ on its own interface.
On the contrary, the third term arises from the interaction potential and describes a force induced by the surrounding cells on the interface of cell $i$.

These expressions can be used to obtain a sensible definition of the tissue pressure as follows.
We require the surface tension and compression terms to induce a force on the neighbouring cells $k \neq i$ as well, even though such terms do not appear in $\delta F / \delta \phi_i$.
This is reasonable because we expect compressed or highly elongated cells to exert a restoring force on their neighbours proportional to the degree of their compression.
Because the corresponding gradients $\nabla \phi_k$ are pointing in the opposite direction, it means that we must invert the sign in front of these terms.
On the contrary the forces arising from the interaction already have the correct sign.
This leads to the definition
\[
P = \sum_{i}\left( \frac{\delta F_{\text{rep.}}}{\delta \phi_i} - \frac{\delta F_{\text{CH}}}{\delta \phi_i} - \frac{\delta F_{\text{area}}}{\delta \phi_i} \right).
\]
Note that self contributions vanish when integrating over space, which means that terms involving only the phase field of a given cell $\phi_i$ do not contribute to $\vect F_i$.
This definition can be easily shown to be thermodynamically consistent by the same argument as the one presented in \cite{Palmieri15}.

\subsection{Deformation tensor of single cells}
The deformation of single cells can be characterised by considering the structure tensor of their phase field, see for example~\cite{doi:10.1007/s10035-003-0127-9}.
We define the deformation tensor $\matr S_i \equiv \matr S (\phi_i)$, with
\[
\matr S(\phi)
= \begin{pmatrix} S_{11} & S_{12} \\ S_{12} & - S_{11} \end{pmatrix}
= \int \dd \vect x \,
\begin{pmatrix}
\frac 1 2 \left( (\partial_y \phi )^2 - (\partial_x \phi )^2 \right) & -(\partial_x \phi ) (\partial_y \phi) \\
- (\partial_x \phi ) (\partial_y \phi)  & \frac 1 2 \left( (\partial_x \phi )^2 - (\partial_y \phi )^2 \right) 
\end{pmatrix},
\]
which is the traceless part of the negative of the structure tensor $-\int \dd \vect x\, (\nabla \phi)^\transp \nabla \phi$.
Adding a negative sign is necessary here because the elongation axis corresponds to the direction of smallest gradients since the strength of these gradients is approximately independent of the shape of the interface.
The eigenvalues and eigenvectors of $\matr S$ capture the deformation of the cell.
Defining $\mathrm S v^{\pm} = \lambda^{\pm} v^{\pm}$, they are explicitly given by
\begin{align*}
\lambda^\pm &= \pm r, \\
\vect v^+ &= (\cos(\omega),   \sin(\omega)), \\
\vect v^- &= (\sin(\omega), - \cos(\omega)),
\end{align*}
with  $r = \sqrt{S_{11}^2 + S_{12}^2 }$ and $\omega = \tan^{-1}(S_{11}/S_{12})/2$.
The magnitude $r$ gives the strength of the deformation, while the eigenvectors $\vect v^+$ and $\vect v^-$ define the orientations of largest elongation and contraction, respectively.
Note that the eigenvectors are defined up to a multiplicative constant, such that they do not define a direction but only a deformation axis.
Connection to the usual representation of a nematic tensor in terms of its director in two dimensions can be easily obtained by solving $S_{11}$ and $S_{12}$ for $r$ and $\omega$, giving
\[
\matr S = 2 r (\vect n \vect n^\transp - \vect n^2 \mathbb I/2) = r \begin{pmatrix} \cos 2 \omega & \sin 2 \omega \\ \sin 2 \omega & - \cos 2 \omega \end{pmatrix},
\qquad \text{with} \quad
\vect n = \vect v^+.
\]
That is, the deformation tensor $\matr S$ corresponds to a nematic tensor with director $\vect v^+$ and order parameter $r$.


\subsection{Scaling properties of the activity threshold}

The linear dependence of the critical activity on the elasticity shown on Fig.~\ref{fig:transition}(b) can be qualitatively explained by considering the balance of stresses in Eq.~\ref{eq:zeta}, in the main text.
    Here, we conjecture that the activity threshold is set by the activity strength required to deform a cell, i.e. it is given by the magnitude of the active stress $\zeta\matr{Q}=\zeta\Sigma_i\phi_i\matr{S}_i$ (Eq.~\ref{eq:Q}) needed to overcome the pressure due to the elastic energetic cost coming from the Cahn-Hilliard contribution $\Sigma_i\delta\mathcal{F}_\text{CH}/\delta\phi_i$ (Eq.~\ref{eq:pressure}).
    Using definitions of the shape tensor $\matr{S}$ and the Cahn-Hilliard free energy $\mathcal{F}_\text{CH}$, we can write $\matr{S}_i\sim\int d\vect{x}(\vect{\nabla}\phi_i)^\text{T}\vect{\nabla\phi_i}$ and $\delta\mathcal{F}_\text{CH}/\delta\phi_i\sim\int d\vect{x}\gamma\lambda(\nabla^2\phi_i)$. Balancing these two and noting that the gradients in phase fields occur over the interface width (i.e., $\vect{\nabla}\phi\sim 1/\lambda$) and are therefore independent of $\gamma$ or $\zeta$, a simple scaling argument yields $\zeta_{\text{cr}}\sim\gamma$ and hence the linear dependence on the elasticity $\gamma$.

\subsection{Fast simulation method using domain decomposition}

Compared to models where the individual cells are described using a tesselation of the plane~\cite{doi:10.1103/PhysRevX.6.021011, doi:10.1371/journal.pcbi.1005569}, the phase field approach has the advantages that it can accommodate large shape deformations of the cells independently of the location of their center-of-masses and that it naturally allows for the description of tissue boundaries.
Its main disadvantage has been the high computational cost associated with describing each cell with a different phase field, but we present here a novel computational algorithm that makes the efficient simulation of large epithelia possible.

Solving Eqs.~(1--4) of the main text efficiently is made difficult by our choice of using a different phase field for each cell. 
In particular both the computational and memory requirements grow quadratically with the number of cells $N$.
To be more precise, let us consider a confluent epithelium in a domain of area $A$ and let us fix the size $R$ of the cells.
In order for the epithelium to stay confluent while increasing $N$, the domain area must scale as $A \sim N$ to accommodate the new cells.
Because we simulate each cell as a separate phase field defined on the whole domain, it follows that the total computational and memory requirements must scale as $NA \sim N^2$.
This makes the simulation of systems comprising a large number of cells impractical.

We solve this problem by restricting each field to a subdomain centred on its centre of mass.
Because the equilibrium profile of the phase fields is exponentially falling, it can be neglected far from its centre of mass.
The centre-of-mass $\vect c(\phi) = (c_x(\phi), c_y(\phi))$ of a phase field $\phi$ in the presence of periodic boundaries can be obtained as
\begin{equation}
\label{eq:com}
c_s(\phi) = \frac {L_s} {2\pi} \arg\left( \int_0^{L_x} \dd x \int_0^{L_y} \dd y\,\, \phi (x, y)\, \mathrm e^{2 \pi i s/L_s} \right),
\end{equation}
where $s=x, y$ is one of the coordinates, $L_x$ and $L_y$ are the domain lengths and $\arg$ is the complex argument defined on $[0, 2 \pi]$.
This amounts to computing the centre-of-mass in Fourier space and transforming back such that periodic boundary conditions are taken into account automatically.
We then restrict each phase field to be non-zero only in a fixed-size subdomain around its centre of mass and only store these values in memory.
These values can then be mapped onto the full domain using the location of the corresponding centre of mass.

\begin{figure*}[h!]
    \centering
    \raisebox{3.7cm}{a}\hspace{.2cm}
    \begin{picture}(230,100)
    \put(0,0){\includegraphics[height=4cm]{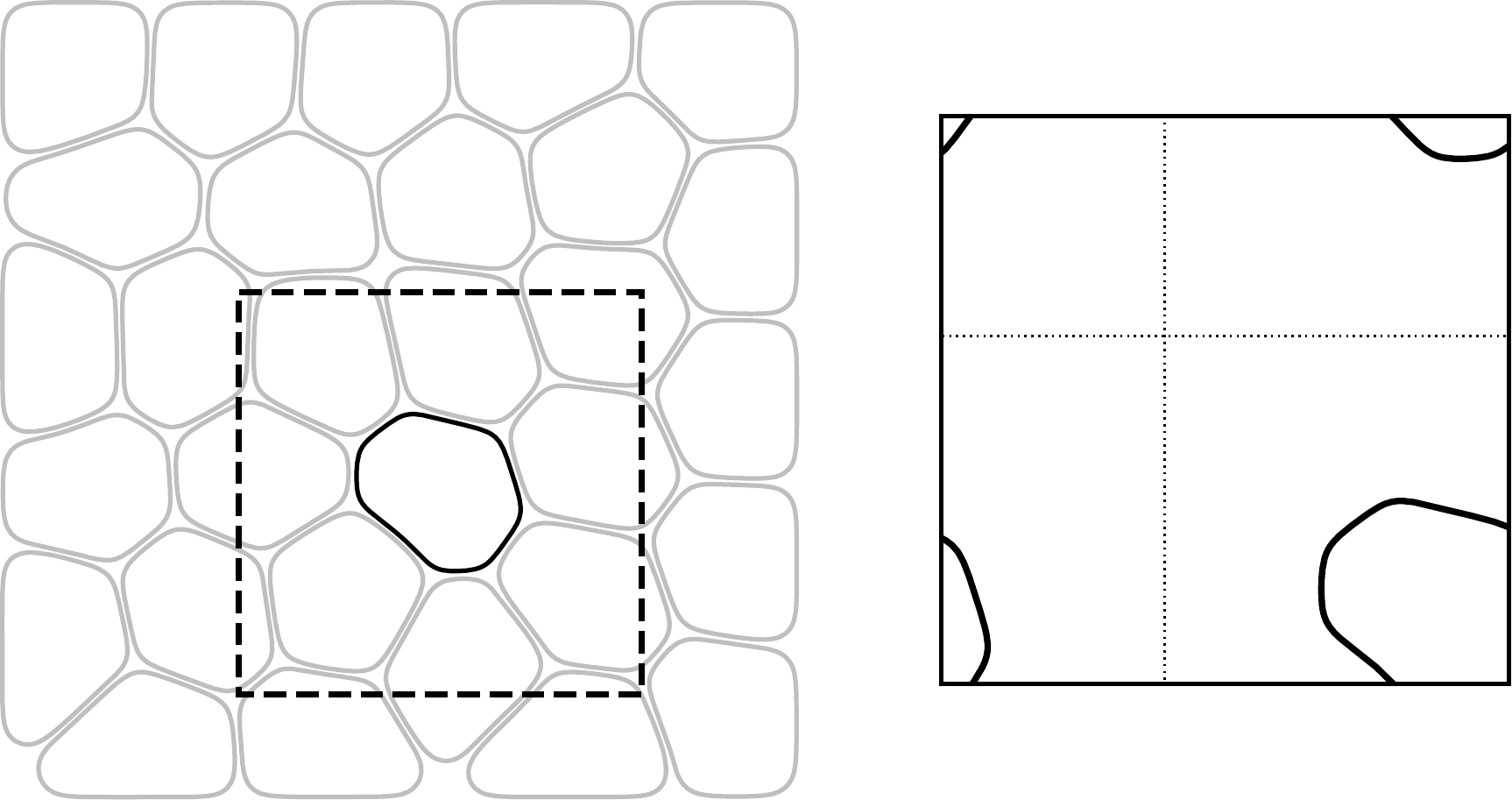}}
    \put(31,120){Full domain}
    \put(152,120){Subdomain}
    \end{picture}
    \raisebox{4cm}{b}\includegraphics[height=4.2cm]{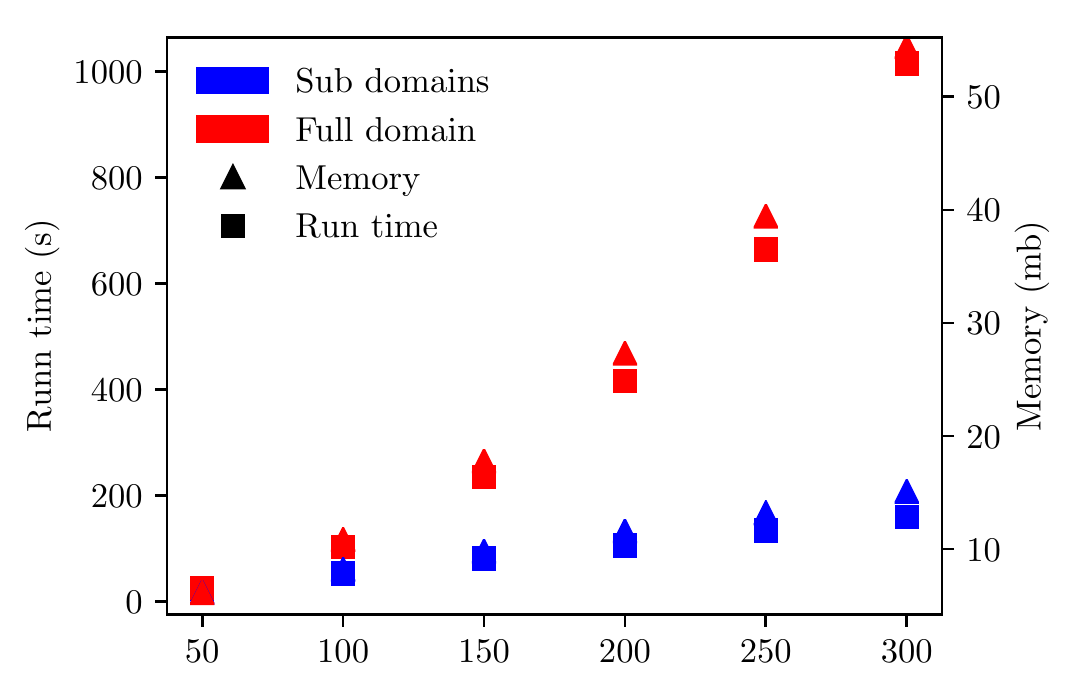}
    \caption{\label{fig:implementation}
        (a), Subdomain associated with a single cell (left) and the corresponding memory layout (right) with dynamical offset shown as dotted lines.
        (b), Runtime and memory usage of the subdomain algorithm compared to simulating the full box for different domain sizes $L$. We simulated confluent monolayers of sizes $50\times L$ with a cell density of 14 cells per $50 \times 50$ lattice sites.
        Simulations performed on a Intel Core i7-7600U CPU (single threaded).
    }
\end{figure*}

However, this approach presents two distinct problems.
First, because the centre of mass can lie between lattice points, this mapping changes abruptly every time the nearest lattice point to the centre of mass changes.
Second, it is then necessary to shift all values of the phase field in the subdomain in order to preserve the location of the cell in the full domain, resulting in costly copying operations.
We solve both of these problems by using \emph{periodic boundary conditions on the subdomain} and by introducing a two-dimensional offset that tracks the upper-left corner of the subdomain.
This allows us to shift the offset every time the subdomain hops without inducing any artefact in the phase field of the cell enclosed and without using any copying operation.
An illustration of this algorithm is shown in Suppl. Fig.~\ref{fig:implementation}a.

This technique can be used to integrate our system of equations efficiently.
Since we only simulate cells in their respective subdomain, both the computational and memory costs of the algorithm now scale as $\sim N A_{\text{subdomain}} \sim N$ for a confluent epithelium.
Note that even with this restriction, it is still necessary to compute some quantities on the whole domain, such as the sum of all phase fields $\sum_i \phi_i$, but the number of such quantities is independent of the total number of cells.
We illustrate how the computational time and memory requirements scales with increasing epithelium size in Suppl. Fig.~\ref{fig:implementation}(b).
This method has been found to give satisfactory results and to agree qualitatively with the naive simulation of Eq.~1 in the main text.
Another recently proposed approach that requires much more theoretical work is to consider the sharp interface limit of the model and simulate the interfaces explicitly~\cite{arXiv:1807.07836}.

\subsection{Implementation details}

We simulate equations~(1--4) in the main text using a finite-difference scheme on a square lattice with a predictor-corrector step.
Thanks to the algorithm described in the previous section, we consider only nodes that are included in a subdomain associated with a cell and are able to achieve linear computational and memory complexity with respect to the number of cells, see Suppl. Fig.~\ref{fig:implementation}(b).
We implemented our algorithm in \texttt{C++} and parallelised it to multi-core architectures using \texttt{OpenMP} and to GPU using \texttt{Cuda}.
Note that our algorithm is rather efficient and allows the simulation of epithelia of about thousand cells in a matter of hours on a personal computer without using GPU acceleration.
We have observed that GPU acceleration improves this baseline by about a factor of 100 using a GeForce GTX 1080 Ti.

All the double sums appearing in $\delta \mathcal{F} / \delta \phi_i$ can be dealt with efficiently by storing global fields such as $\sum_i \phi_i$ and $\sum_i \phi_i^2$ at every point of the domain.
Because the total force $\vect F_i^{\text{tot.}}$, involves integrals over the whole domain we use a simple two-pass algorithm where the phase fields $\phi_i$ are updated one after another.

Unless otherwise stated, simulation parameters are $\gamma  = 0.01$, $\mu = 3$, $\lambda  = 2.5$, $\kappa = 0.1$, $R = 8$, $\xi = 2$ and we use a subdomain size of $25 \times 25$ lattice sites.
Comparing these parameters to the average radius of MDCK cells $\sim 5\mu$m, the typical velocity $\sim 20\mu$m/h and the pressure $\sim 100 \text{Pa}$ measured in MDCK monolayers using Particle Image Velocimetry and Traction Force Microscopy~\mbox{
        \cite{Saw17}}\hspace{0pt}
    , we find $\Delta x\sim0.5\mu$m, $\Delta t\sim0.1$s, and $\Delta F\sim 1.5\text{nN}$ as the space, time, and force corresponding to simulation units.

Simulations of different sizes have been done at constant density of $15$ cell per $50\times 50$ lattice points. This corresponds for example to $60$ cells for domains of size $100\times100$ and to $375$ cells for domains of size $250\times 250$. 
All simulations are performed with periodic boundary conditions and the system is initialised by creating cells at random positions.
Cells initially have a smaller area than their target area of $\pi R^2$ and we allow them to relax without activity for some initial time.
We store the full information of each phase field $\phi_i$ on their subdomain and perform all the post-processing and plotting using \texttt{Python}.

\begin{figure}[t]
    \includegraphics[width=2in]{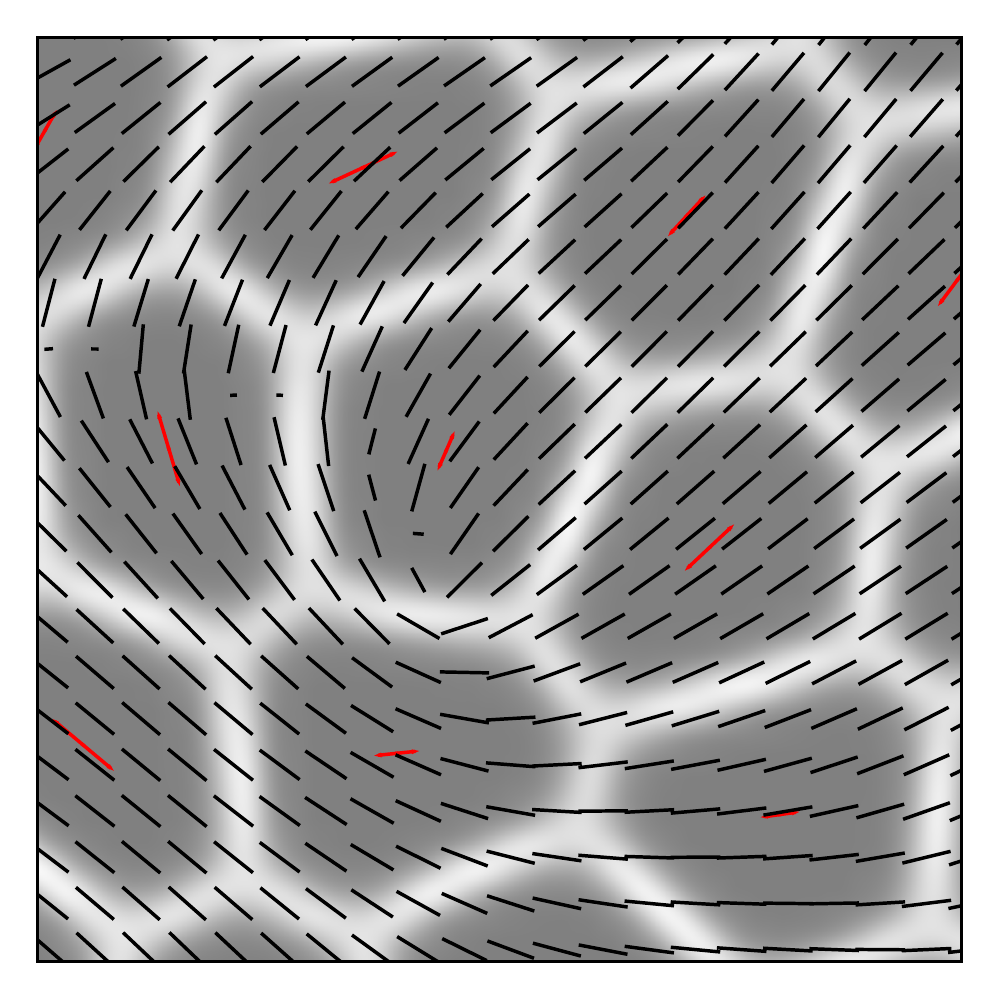}\\
    \caption{\label{fig:zoom defect} %
        Region of size $40\times 40$ cropped around a single positive defect. The deformation field $\matr Q$ (shown in black) has been smoothed using a sliding window of size $3R\times 3R$ and is only showed every second lattice point. The directors of the individual shape tensors of each cell are shown in red.}
\end{figure}

Local properties of flows and stresses around defects are obtained by first smoothing the global director field $\matr Q$ using a sliding window of size $3R \times 3R$ in order to reduce the noise, see Suppl. Fig.~\ref{fig:zoom defect}.
    Note that because each deformation tensor $\matr S_i$ is weighted with the corresponding phase field $\phi_i$, the deformation field $\matr Q$ is defined over all space even before smoothing.
    We then track all defects using a custom tracking algorithm and obtain their orientation by following the method presented in reference~\mbox{
        \cite{Vromans16}}\hspace{0pt}
    .
    We then crop and align the fields around each defect and average over all defects as well as over the whole time of the simulations.
    All of the simulations are started with random initial conditions and the quantitative results are averaged over at least 5 different random initialisations.

\clearpage
\section{Supplementary figures}

\begin{figure}[h!]
    \includegraphics[width=3in]{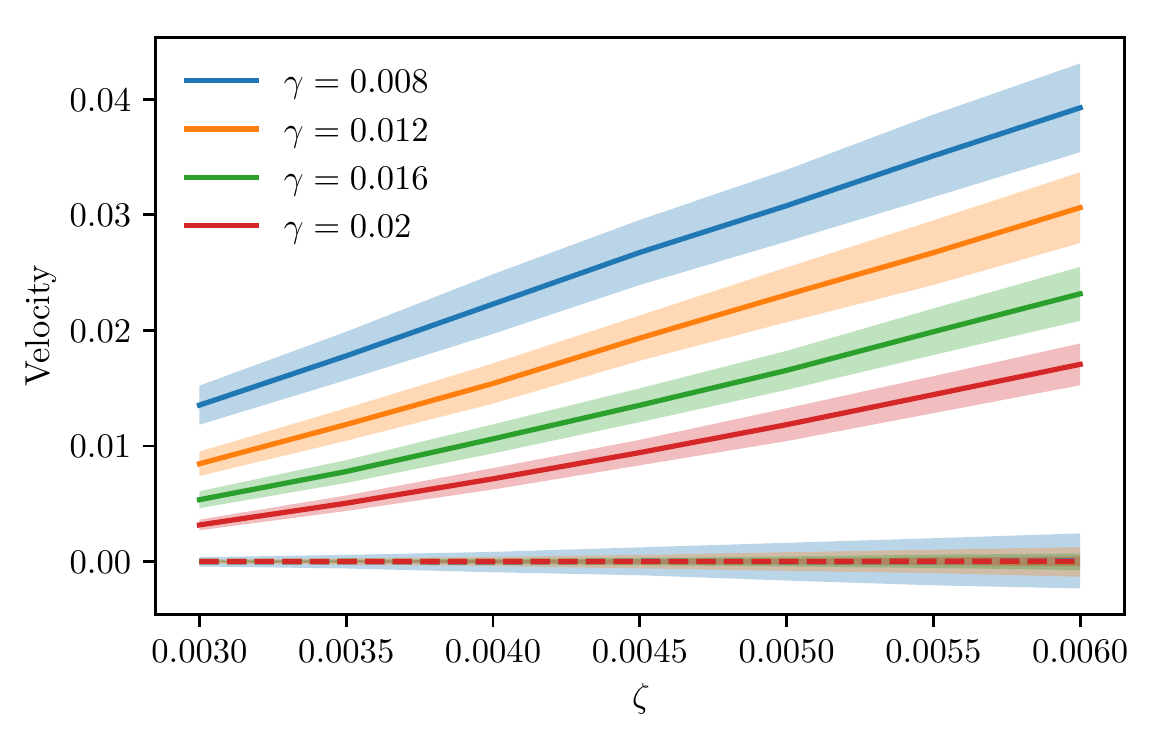}
    \caption{%
        \label{fig:velocity}
        Root-mean-square velocity $v^2_{\mathrm{rms}} = \langle \vect v^2 \rangle$ (plain lines) and square of the mean velocity $\langle \vect v \rangle^2$ (dashed lines) where averages are computed over the full velocity field smoothed over three cell radii.
        Mean$\pm$std from $5$ simulations.
    }
\end{figure}

\begin{figure}[h!]
    \includegraphics[trim={0 0 0 0},clip,width=0.5\linewidth]{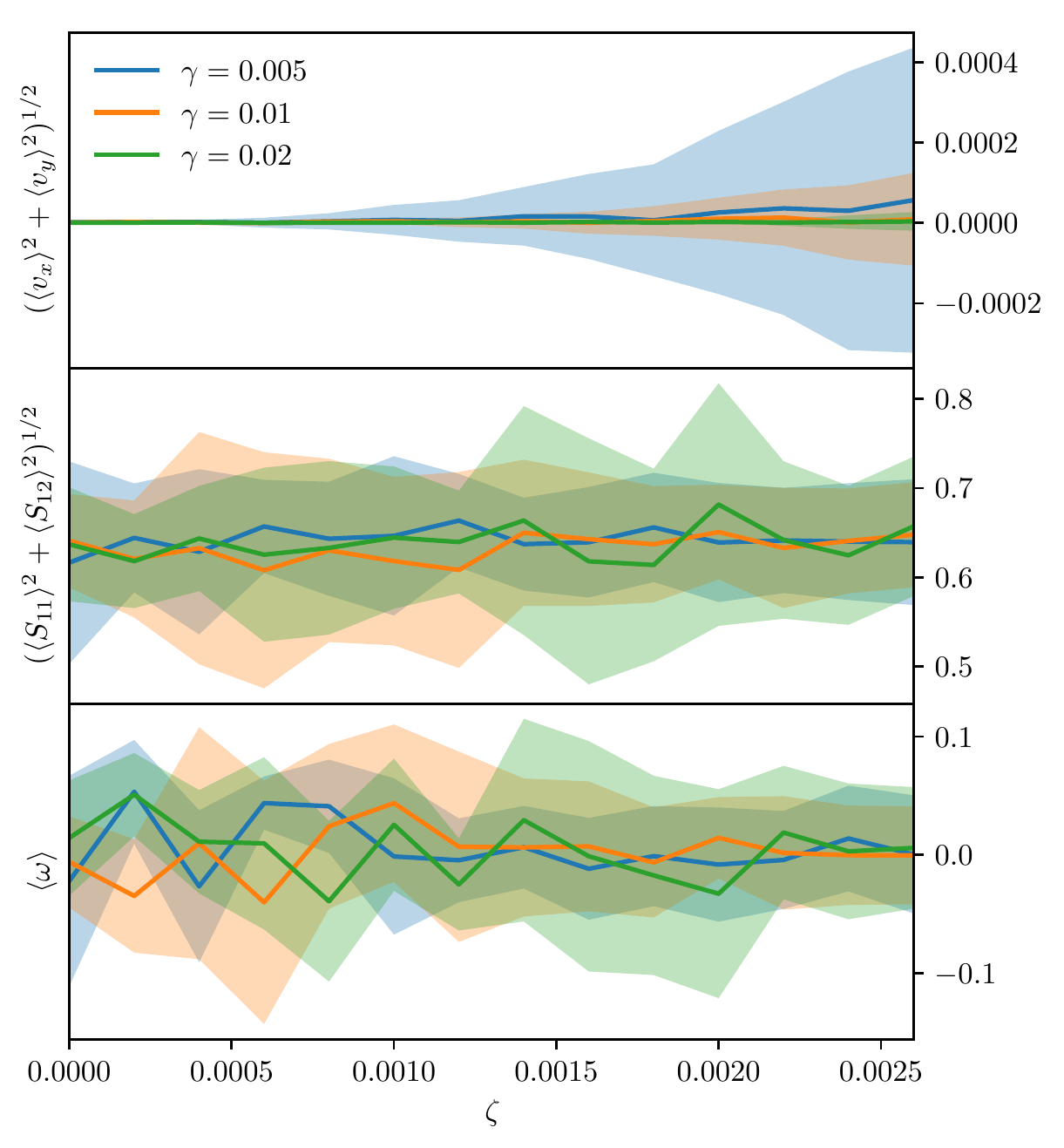}
    \caption{%
        \label{fig:transition-check}
        Square roots of the sums of the individual components of the velocity $\vect v$ and order $\matr S$ do not show any noticeable increase in mean value with increasing activity indicating Galilean invariance of the transition to turbulent flows.
        The mean angle of the director $\omega = \text{atan}(S_{12}/S_{11})$ is uniformly distributed in $[-\pi, \pi]$ such that its mean value vanishes.
        This shows that even if cells have a residual order due to deformations at close packing this does not correspond to any alignment of the director.
    }
\end{figure}

\end{document}